\documentclass[fleqn,twoside]{article}
\usepackage{espcrc2}
\usepackage[english]{babel}
\usepackage{amssymb}
\usepackage{latexsym}
\usepackage{graphicx}
\usepackage[dvips]{epsfig}

\title{Final QCD results from LEP}

\author{H.\ Stenzel\address{II. Physikalisches Institut, Universit\"at Giessen, 
        Heinrich-Buff Ring 16, D-35392 Giessen, Germany}
\thanks{Contribution to the Proceedings of 'QCD 04 International 
                Conference in Quantum Chromodynamics', Montpellier, France, 
                July 5 - 10, 2004}}

\begin{document}
\newcommand{\unit}[1]{\,{\mathrm{#1}}}
\newcommand{\as}{\overline{\alpha}_s}
\newcommand{\epem}   {\ensuremath{\mathrm{e^+e^-}}}
\newcommand{\roots} {\sqrt{s}}
\newcommand{\ecm} {\ensuremath{E_{\rm cm}}}
\newcommand{\mz}     {\ensuremath{M_{\rm Z}}}
\newcommand{\mw}     {\ensuremath{M_{\rm W}}}
\newcommand{\bm}[1]  {\mbox{\boldmath $#1$}}
\newcommand{\bt}     {\ensuremath{B_T}}
\newcommand{\bn}     {\ensuremath{B_N}}
\newcommand{\bw}     {\ensuremath{B_W}}
\newcommand{\mh}     {\ensuremath{M_h^2/s}}
\newcommand{\mhd}    {\ensuremath{(M_h^2-M_l^2)/s}}
\newcommand{\thr}    {\ensuremath{1-T}}
\newcommand{\ptin}    {\ensuremath{p_{\perp}^{\rm in}}}
\newcommand{\ptout}    {\ensuremath{p_{\perp}^{\rm out}}}
\newcommand{\nch}    {\ensuremath{\langle N_\mathrm{ch}\rangle}}
\newcommand{\tmaj}   {\ensuremath{T_{\mathrm{major}}}}
\newcommand{\tmin}   {\ensuremath{T_{\mathrm{minor}}}}
\newcommand{\oaa}    {\ensuremath{\mathcal{O}(\alpha_s^2)}}
\newcommand{\alphab} {{\overline \alpha_s}}
\newcommand{\xp}     {\ensuremath{x_p}}
\newcommand{\xmu}     {\ensuremath{x_\mu}}
\newcommand{\mui}     {\ensuremath{\mu_{\rm I}}}
\newcommand{\ksip}   {\ensuremath{\xi}}
\newcommand{\klphd}   {\ensuremath{K_{\rm LPHD}}}
\newcommand{\xistar} {\ensuremath{\xi^*}}
\def\half{\mbox{\small $\frac{1}{2}$}}
\newcommand{\la}{\langle}
\newcommand{\ra}{\rangle}
\newcommand{\myn}    {\ensuremath{\langle y^n\rangle}}
\newcommand{\vecma}  {\ensuremath{\vec{n}_{\mathrm{Ma}}}}
\newcommand{\vecmi}  {\ensuremath{\vec{n}_{\mathrm{Mi}}}}
\newcommand{\vecsma}  {\ensuremath{\vec{n}_{\mathrm{sMa}}}}
\newcommand{\vecsmi}  {\ensuremath{\vec{n}_{\mathrm{sMi}}}}
\newcommand{\chidof}     {\ensuremath{\chi^2/N_{\rm DOF}}}
\newcommand{\evis}     {\ensuremath{E_{\rm vis}}}
\newcommand{\ycut}     {\ensuremath{y_{\rm cut}}}
\newcommand{\ymax}     {\ensuremath{y_{\rm max}}}
\newcommand{\stot}     {\ensuremath{\sigma_{\rm tot}}}
\def\Covrln{{\overline C}}
\def\Govrln{{\overline G}}
\def\Lovrln{{\overline L}}
\def\Lovrlntilde{{\widetilde{\overline L}}}
\def\Lbar{{\bar L}}
\def\as{\alpha_{s}}
\def\gae{{\gamma_{\textsc{e}}}}
\def\asb{{\bar \alpha}_{{\textsc{s}}}}
\begin{abstract}
  Recent results on QCD studies in $e^+e^-$ annihilations at LEP  
  are presented. Data recorded by the LEP experiments 
  at centre-of-mass energies between 91.2 to 206 GeV are included. The main topic 
  is the measurement of $\alpha_s$ from event shape variables and associated aspects 
  like the energy evolution or non-perturbative power-law corrections. These 'standard' 
  measurements are complemented by new determinations using the 4-jet rate with an 
  excellent precision. A summary of the results on QCD colour factors from  
  angular correlations in the 4-jet system completes this report.   
\end{abstract}

\maketitle

\section{Introduction}
The LEP experiments ALEPH, DELPHI, L3 and OPAL have collected 700 pb$^-1$ of 
annihilation data from 1989 to 2000. Final QCD analyses of these data using 
the latest detector simulation and advanced correction techniques are now 
being published. The experiments presented cumulative summary papers including 
various centre-of-mass energies and a wide collection of observable. The 
experimental systematic uncertainties are for many observables as small as 1$\%$, 
and the statistical errors typically 0.1 $\%$. Hence, these data of un-preceeded precision 
will serve as reference for future experiments. This report summarises recent measurements of $\alpha_s$ from event shapes and 4-jet observables, investigations of power law corrections 
and results on QCD colour factors.           

\section{Measurements of $\alpha_s$ from event shapes}
Theoretical calculations for event-shape distributions are available at 
next-to-leading order (NLO) complemented by all-orders resummation of large 
leading and sub-leading logarithms (NLLA) for certain observables. A unified prescription 
for the matching of fixed-order and resummed calculations, the so-called 
modified logR matching scheme, has recently been suggested \cite{hasko} 
and is applied  by the LEP experiments for their analysis. 
The most commonly used variables are    
thrust, heavy jet mass, wide and total jet broadenings, C-parameter and $y_3$, the 3-jet 
resolution parameter in Durham scheme. These observables were also selected by 
the LEP QCD working group for a LEP combination \cite{Roger}. A virtue 
of event-shape distributions compared to other observables is that even with 
limited event statistics at LEP2 energies a measurement of $\alpha_s$ is possible 
with a reasonable precision and enables the observation of the energy evolution of 
$\alpha_s$.~  
\begin{figure}[h!]
  \unitlength1mm
  \begin{picture}(75,72)
    \put(-5,-3){\includegraphics{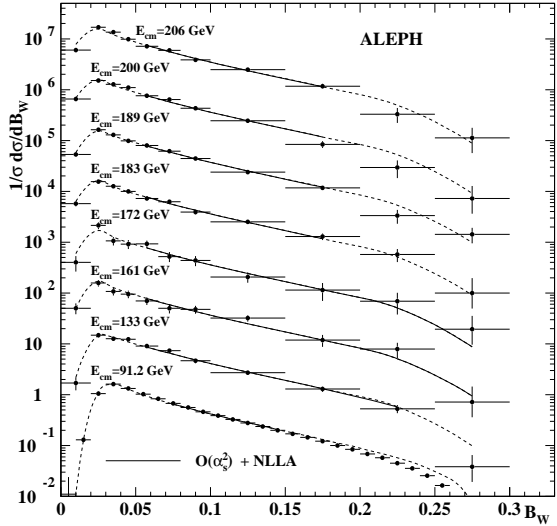}}
  \end{picture}
  \caption{Distributions at various centre-of-mass energies of the wide jet 
  broadening compared to theoretical predictions used to determine $\alpha_s$.}
\label{fig:alephbw}
\end{figure}   
An example of a final analysis is shown in fig.~\ref{fig:alephbw} where 
the measurements of the wide jet broadening variable at all LEP energies are compared to the 
result of fits with the NLO+NLLA theoretical prediction. ALEPH has in addition 
applied a perturbative NLO correction for the b-quark mass. In general the description of the 
data is good, although restricted to the central part of the distributions at LEP1 where 
the high precision of the data requires NNLO calculations. The measurements have been 
corrected for acceptance and detector resolution effects and the fits are carried out 
at hadron level. Therfore, the perturbative prediction is folded to the hadron level 
by means of transition matrix accounting for the hadronisation, obtained from standard 
QCD generators like PYTHIA, HERWIG or ARIADNE. The systematic uncertainties for the 
measurement of $\alpha_s$ is dominated by theoretical uncertainties induced by 
missing higher orders. The LEP QCD working group advises the recommendation of \cite{hasko} 
for the estimation of perturbative uncertainties. This method combines different estimates in 
the 'uncertainty band method', which takes not only standard renormalisation scale variations 
but also a variation of $x_L$, the resummed logarithmic variable re-scaling factor, into account \cite{gavin}.
\subsection{Combined measurements} 
\begin{figure}[h!]
  \unitlength1mm
  \begin{picture}(75,55)
    \put(-2,0){\includegraphics{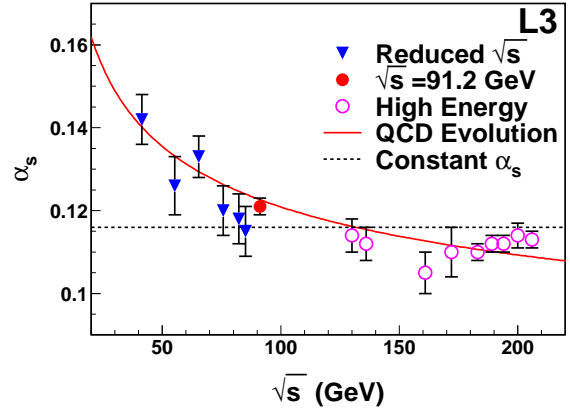}}
  \end{picture}
  \caption{Measurements of $\alpha_s$, combining five event-shape variables, 
  as function of centre-of-mass energy.}
\label{fig:l3run}
\end{figure}

The measurements using different variables and different energies are combined in 
single numbers per $E_{\rm CM}$ and finally using the QCD-predicted evolution in 
global result for $\alpha_s(M_Z)$. 
In fig.~\ref{fig:l3run} the combined result from L3 for $\alpha_s(Q)$ is 
shown, including also measurements using radiative events resulting in reduced centre-of-mass 
energies below $\mz$.~  
The experiments have applied different techniques for 
their combinations in terms of the correlation of systematic uncertainties, but in all cases 
the theoretical uncertainties appear to be largely correlated, both between different variables 
and between different energies. As a consequence, the gain in precision of the combined 
measurements in limited and the combined uncertainty appears as average rather than an improved
uncertainty. 

ALEPH has carried out a global analysis of 6 variables at 8 energies between 
91 and 206 GeV \cite{aleph}, based on an integrated luminosity of about 700 pb$^{-1}$. 
Their combined result for $\alpha_s(M_Z)$ is a weighted average with weights proportional 
to the inverse square of the total individual errors. To account for correlations of systematic 
uncertainties the whole combination is repeated separately for all the individual variations of the 
analysis. The final result is 
$$\alpha_s(M_{\rm Z})=0.1214 \pm 0.0014_{exp} \pm 0.0046_{th}.$$ 

L3 presented in a collective preprint \cite{l3qcd} various studies of QCD, including 
measurements of $\alpha_s$ from an earlier publication \cite{l3as}. They combined the
same five variables, but excluded $-\ln y_3$. Measurements at reduced centre-of-mass energies 
using radiative events were included. Theoretical uncertainties are estimated by a variation 
of the renormalisation scale and different matching schemes. L3 has chosen at LEP2 very 
large fit ranges, in contrast to all other experiments, in order to reduce the statistical 
uncertainty. This leads on the other side to larger theoretical errors. The combination method 
proceeds in two steps: first an unweighted average of the five variables 
is build at each $E_{\rm CM}$, second these combined measurements are fit by the QCD evolution 
for $\alpha_s(M_{\rm Z})$. At that stage correlations are included by the assumption of minimal 
overlap of systematic uncertainties. The result of this combination is:  
$$\alpha_s(M_{\rm Z})=0.1227 \pm 0.0012_{exp} \pm 0.0058_{th}.$$ 

DELPHI updated a publication \cite{delphi} on event-shapes with a new re-analysis \cite{delphinew} 
according to the LEP QCD scheme. Traditionally DELPHI presents measurements based on three perturbative 
methods: the standard modified logR matching scheme, the pure NLLA calculation (valid only in a narrow 2-jet 
region) and the fixed order prediction alone with an optimised scale $x_\mu^{opt}$. The perturbative uncertainties 
are estimated by a variation of $x_L$ only for the logR/NLLA schemes and by a renormalisation scale variation 
of a factor of two around the optimum $x_\mu^{opt}$. The same set of variables as used by L3 is combined at 
energies between 89 and 207 GeV, hence including also off-peak data at 89 and 93 GeV separately. The combination 
technique applied by DELPHI follows closely the LEP QCD Ansatz, but makes in addition the minimum overlap 
assumption for hadronisation and perturbative systematic uncertainties. The final result, split into the three 
perturbative schemes, reads as:
\begin{eqnarray*}
LogR &  & \alpha_s=0.1205 \pm 0.0020_{exp}\pm 0.0050_{th}\; ,\\
{\cal O}(\alpha_s^2) &  & \alpha_s=0.1157 \pm 0.0018_{exp} \pm 0.0027_{th}\; ,\\
NLLA &  & \alpha_s=0.1093 \pm 0.0023_{exp} \pm 0.0051_{th}\; .
\end{eqnarray*}
     
OPAL is currently preparing a new analysis on $\alpha_s$ from event shapes and preliminary 
results were already included in the LEP average. In the last OPAL publication \cite{opaljade} 
data from LEP up to 189 GeV were combined with lower-energy data from JADE at 35 and 44 GeV. 
The variables are the differential 2-jet rate (equivalent to $y_3$) and the mean jet multiplicity 
using the Durham and Cambridge jet-finding algorithms. The combined result is:
$$\alpha_s(M_{\rm Z})=0.1287 \pm 0.0012_{exp} + 0.0034_{th} - 0.0016_{th} ,$$
where a significantly asymmetric uncertainty is observed for the scale variation.  
 
\section{power law corrections}
Non-perturbative effects in hadronic observables
in $e^+e^-$ annihilation are scaling with powers of $1/Q$ and can be described by 
analytical models of power law corrections \cite{power}. 
Power corrections in the spirit of these models are related to infrared
divergences of the perturbative expansion at low scales. Analytical calculations
introduce one additional phenomenological parameter $\alpha_0$,
\begin{displaymath}
\alpha_0(\mui)=\frac{1}{\mui}\int_0^{\mui}\alpha_s(k)dk \; ,
\end{displaymath}
which measures effectively the strength of the coupling up to an infrared
matching scale $\mui$ of the order of a few GeV. The parameter $\alpha_0$ is
expected to be universal and must be determined by experiment, usually in conjunction 
with $\alpha_s$.
 
\begin{figure}[h!]
  \unitlength1mm
  \begin{picture}(75,108)
    \put(0,0){\includegraphics{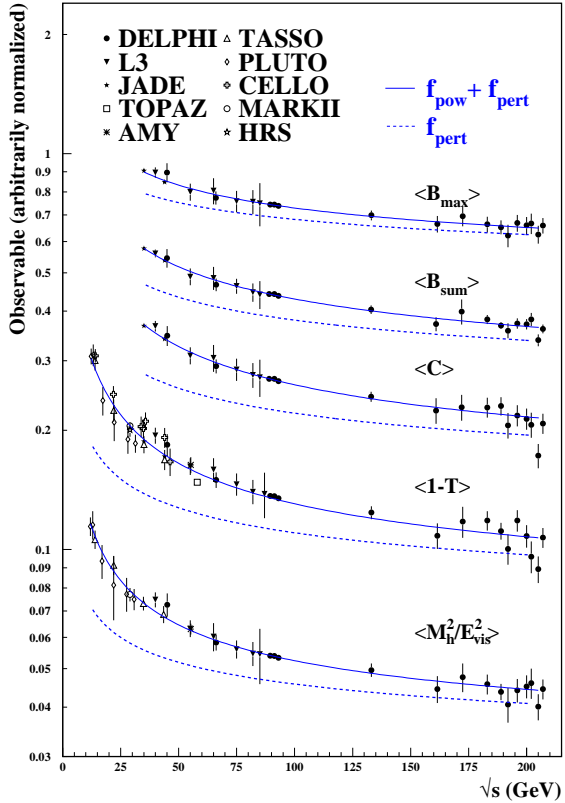}}
  \end{picture}
  \caption{Measurements of event-shape mean values 
  as function of centre-of-mass energy compared to fits including power corrections.}
\label{fig:pcmean}
\end{figure}

An improved theoretical prediction is obtained by merging perturbative and non-perturbative 
terms. This yields for event-shape mean values, 
$$ \langle y \rangle  =  \langle y_{\rm pert} \rangle + \langle y_{\rm power} \rangle ,$$
for a generic variable $y$, where the additive power correction term is given by  
$$\langle y_{\rm power}\rangle  =  c_y \cdot P(\alpha_0)/Q,$$
with a variable-dependent constant $c_y$.
In event-shape distributions the power correction appears as a shift of the 
perturbative spectrum by the same additive term
$$ D_y(y)  =  D_{\rm pert}\left( y- \frac{c_y\cdot P(\alpha_0)}{Q} \right).$$ 
For the jet broadenings this shift is not constant but depends on the value 
of the broadening. In fig.~\ref{fig:pcmean} different mean values are shown as function 
of $\sqrt{s}$ and compared to pure perturbative and power-corrected predictions.~
\begin{table}[h!]
\caption[power correction]{\label{tab:pc}{\small Combined results
on $\alpha_s(\mz.)$ and $\alpha_0$(2 GeV) using different variables. 
Partly mean values or distributions were analysed.}}
\begin{tabular}{lll}
    &  $\alpha_s(\mz)$ & $\alpha_0$(2 GeV) \\ \hline
ALEPH & 0.1112 $\pm$ 0.0053 & 0.496 $\pm$ 0.101 \\
distr. & & \\ \hline
L3 &  0.1126 $\pm$ 0.0060 &  0.478 $\pm$ 0.059  \\
means & & \\ \hline
DELPHI &  0.1110 $\pm$ 0.0055 & 0.559 $\pm$ 0.073 \\
distr. & & \\ \hline
Movilla et al. &  0.1171 $\pm$ 0.0026 & 0.513 $\pm$ 0.050  \\
distr.+means & & \\ \hline
\end{tabular}
\end{table}

\begin{figure}[h!]
  \unitlength1mm
  \begin{picture}(75,95)
    \put(-5,-1){\includegraphics{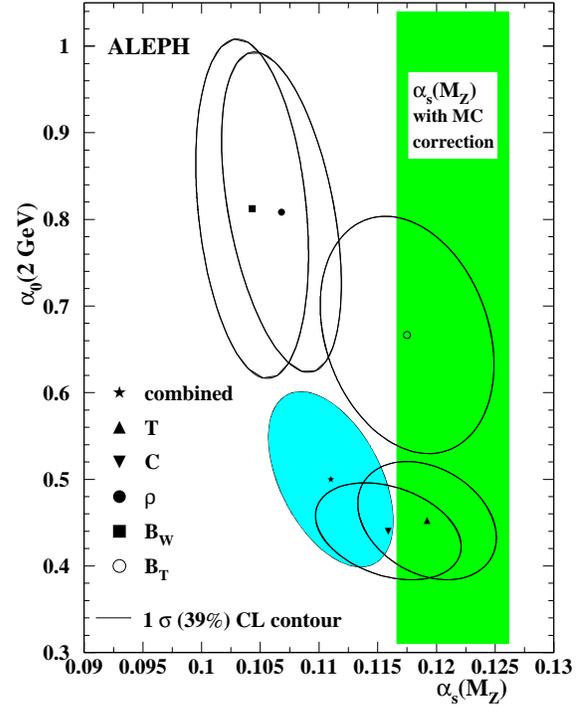}}
  \end{picture}
  \caption{Contours of confidence level for simultaneous
measurements of $\alpha_s$ and $\alpha_0$ (ellipses) compared to
the combined measurement of $\alpha_s$ using Monte Carlo
corrections (shaded band).}
\label{fig:pc_ellips}
\end{figure}

\begin{figure}[h!]
  \unitlength1mm
  \begin{picture}(75,95)
    \put(-5,-1){\includegraphics{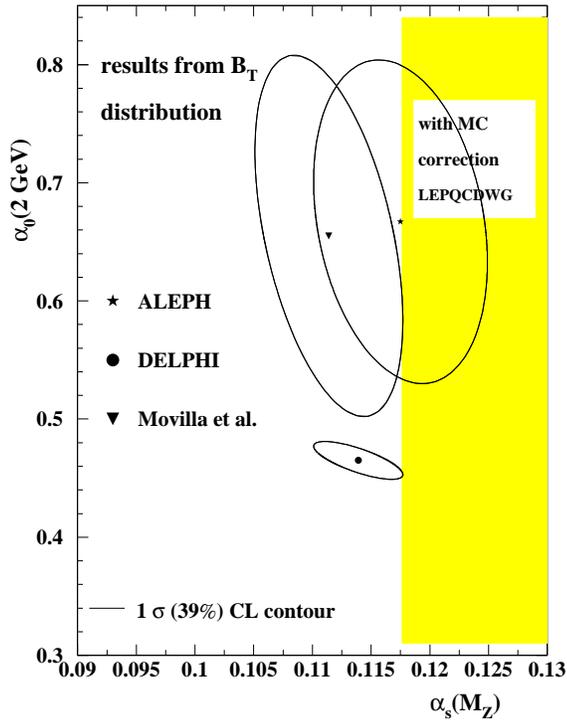}}
  \end{picture}
  \caption{Comparison of confidence level contours from different groups 
for determinations of $\alpha_s$ and $\alpha_0$ using the total jet broadening.}
\label{fig:pc_univ}
\end{figure}

A good description of the data is achieved both for mean values and the central 
part of distributions. Different groups have analysed often similar datasets including 
lower energy measurements and determined the two parameters $\alpha_s(\mz)$ and $\alpha_0$(2 GeV) 
from a simultaneous fit. 
The combined final results are given in table \ref{tab:pc}. The value of $\alpha_0$(2 GeV) 
is close to $0.5$ and the value of $\alpha_s(\mz)$ around $0.114$, significantly lower 
than with standard Monte Carlo corrections for hadronisation. The two parameters in the simultaneous 
fit are strongly correlated. It is instructive to consider the results in the plane $\alpha_s(\mz)$
 versus $\alpha_0$(2 GeV), shown for the ALEPH measurements in fig.~\ref{fig:pc_ellips}.    
It appears that the jet broadening variables prefer a significantly higher value of $\alpha_0$ 
and a lower value of $\alpha_s$, incompatible with thrust and C-parameter, and with $\alpha_s$ 
determined with Monte Carlo corrections. The large systematic uncertainty for $\alpha_0$ from 
the jet broadenings is traced back to uncertainties of the perturbatively calculated power 
correction term, which might be responsible for the large spread between different variables,  
spoiling the predicted universality of $\alpha_0$. 
Furthermore the findings of different groups are not always consistent with each other, as 
demonstrated in fig.~\ref{fig:pc_univ}, where central values and contour ellipses of 
ALEPH and Movilla et al. are consistent while the value of $\alpha_0$ from DELPHI is 
much lower and the error ellipse significantly smaller.  
 


\section{4-jet observables}
Traditionally angular 4-jet observables have been used to determine QCD color 
factors, often in conjunction with a determination of $\alpha_s$ from the 
4-jet rate. The precision of the measurements is substantially improved 
with the advent of NLO $\cal O$$(\alpha_s^3)$ calculations. Two analyses 
have been carried out recently by the ALEPH \cite{aleph_4jets} and OPAL \cite{opal_cacf} 
collaborations, 
which determined simultaneously the colour factor ratios $T_R/C_F$, 
$C_A/C_F$ and the coupling constant. 
Using the QCD value of normalisation $T_R$=$1/2$, ALEPH obtained:      
\begin{eqnarray*}
C_A & = & 2.93 \pm 0.14_{stat} \pm 0.58_{sys}\; , \\
C_F & = & 1.35 \pm 0.07_{stat} \pm 0.26_{sys}\; , \\
\alpha_s(\mz) & = & 0.119 \pm 0.006_{stat} \pm 0.026_{sys}\; .
\end{eqnarray*}
These measurements of the colour factors are excellent agreement with 
the QCD expectations $C_A=3$ and $C_F=4/3$. The OPAL analysis yields
\begin{eqnarray*}
C_A & = & 3.02 \pm 0.25_{stat} \pm 0.49_{sys}\; , \\
C_F & = & 1.34 \pm 0.13_{stat} \pm 0.22_{sys}\; , \\
\alpha_s(\mz) & = & 0.120 \pm 0.011_{stat} \pm 0.020_{sys}\; ,
\end{eqnarray*}
again in good agreement with the ALEPH measurement.
\begin{figure}[h!]
  \unitlength1mm
  \begin{picture}(75,82)
    \put(0,-5){\includegraphics{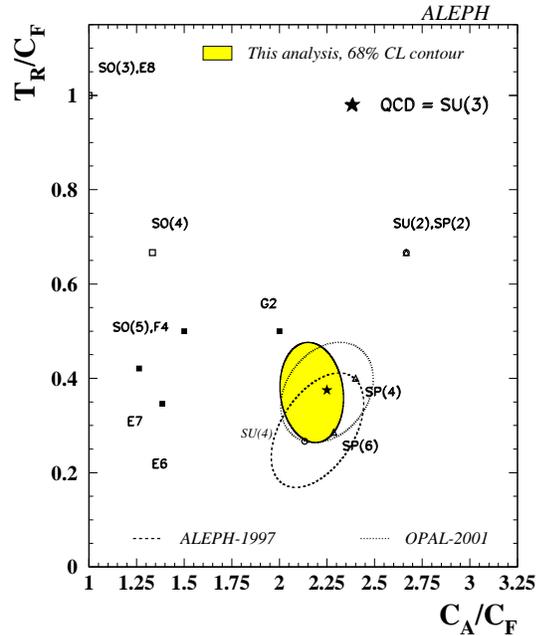}}
  \end{picture}
  \caption{Measurements of the colour factor ratios compared to various gauge group  
  values.}
\label{fig:aleph_su3}
\end{figure}   
The confidence level contours of 
the ALEPH and OPAL analyses are shown in fig.~\ref{fig:aleph_su3} in the plane $T_R/C_F$ versus
$C_A/C_F$. The measurements confirm the QCD expectation of SU(3).
\subsection{$\alpha_s$ from the 4-jet rate}
Alternatively, setting the colour factors to QCD values, $\alpha_s$ can be determined from 
certain 4-jet observables like the 4-jet fraction at NLO. A better sensitivity is 
obtained compared to 3-jet observables, since the leading term is already in $\alpha_s^2$. For 
the 4-jet rate defined in the Durham and Cambridge schemes also resummed calculations are available, 
although only for the R matching scheme. 
ALEPH \cite{aleph_4jets} and DELPHI \cite{delphi_4jets} have presented  measurements 
of $\alpha_s$ using this technique, a new OPAL analysis including also data at LEP2 and from JADE  
is under way.      

\begin{figure}[hb!]
  \unitlength1mm
  \begin{picture}(75,73)
    \put(-1,-4){\includegraphics{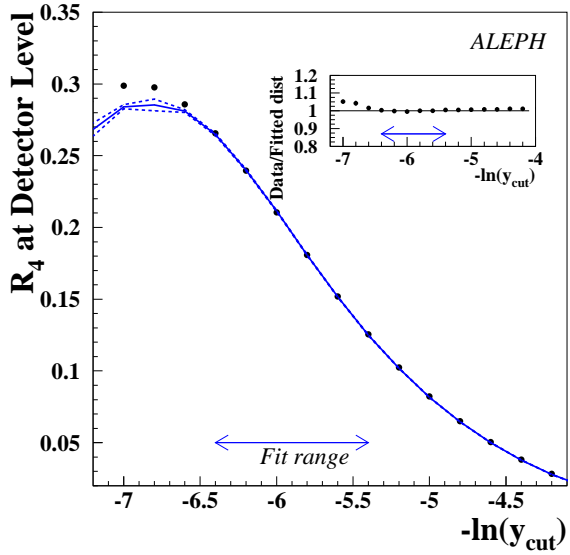}}
  \end{picture}
  \caption{The 4-jet fraction at detector level as function of $\ln y_{\rm cut}$ compared
to a fit of the $\cal O$$(\alpha_s^3)$+NLLA prediction.}
\label{fig:aleph_r4}
\end{figure}   
ALEPH measured the 4-jet rate in the Durham at LEP1 only and determined $\alpha_s$ from 
a fit of the $\cal O$$(\alpha_s^3)$+NLLA (R matching scheme) prediction, as shown in fig.~\ref{fig:aleph_r4}. 
The data are well described inside the fit range provided that the renormalisation scale is 
set the experimentally optimised value $x_\mu^{opt}=0.73$. The quality of the fit becomes rapidly 
worse for scales off the optimum. Systematic uncertainties for the 4-jet analysis are not dominated 
by perturbative but hadronisation uncertainties. This is related to the fact that 4- and 5-jet production 
are less well described in standard Monte Carlo generators. ALEPH applied a Bayesian method to determine 
the size of systematic uncertainties, which consists of a de-weighting of models or theories yielding 
a bad description of the data (large $\chi^2$) by re-scaling the resulting uncertainties. This procedures 
leads to very small errors. 

\begin{figure}[hb!]
  \unitlength1mm
  \begin{picture}(75,100)
    \put(-95,-95){\includegraphics{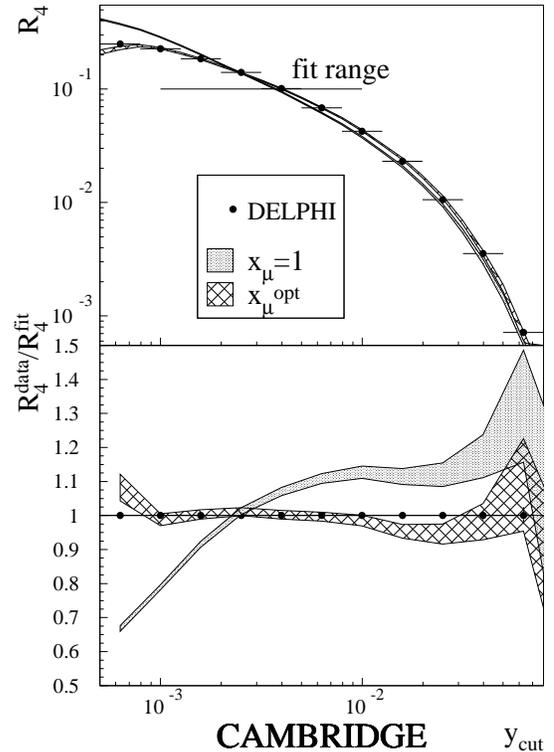}}
  \end{picture}
  \caption{Fits to the 4-jet rate measured at LEP1 for two different scale evaluation methods.}
\label{fig:delphir4}
\end{figure} 
DELPHI investigated the two jet finding schemes Durham and Cambridge and concluded that 
the Cambridge scheme has smaller systematic uncertainties. Their measurement and fit result are 
shown in fig.~\ref{fig:delphir4}, where also the ratio of data over theory is given for $x_\mu=1$ and 
$x_\mu^{opt}$.      
DELPHI used in contrast to ALEPH only the fixed order calculation and found a need for 
rather small scales, $x_\mu^{opt}=0.015$ for Durham and $x_\mu^{opt}=0.015$ for Cambridge. 
The perturbative systematic uncertainty was estimated by a variation of a factor of 2 around 
$x_\mu^{opt}$, this gives rise to very small uncertainties. DELPHI confirms that the main systematic 
uncertainty is associated with the description of $R_4$ in the QCD generators.  
The 4-jet measurements are summarised in table \ref{tab:r4}, the ALEPH result is 
for a direct comparison with DELPHI also given in a non-Bayesian approach for the systematic error. 
\begin{table}[h!]
\caption[4-jets]{\label{tab:r4}{\small Measurements of $\alpha_s$ from the 4-jet fraction.}}
\begin{tabular}{llll}
    & DELPHI & ALEPH & ALEPH \\ 
    & Cambridge & Durham & non-Bayesian \\ \hline
$\alpha_s(\mz)$ & 0.1175 & 0.1170 & 0.1170 \\
exp. & 0.0009 & 0.0008 & 0.0008 \\
had. & 0.0027 & 0.0004 & 0.0021 \\
pert.& 0.0007 & 0.0009 & 0.0016\\
tot. & 0.0030 & 0.0013 & 0.0027\\ \hline 
\end{tabular}
\end{table}
The total uncertainties of about 3 $\%$ for these measurements are very competitive 
compared to methods using 3-jet observables, in particular the perturbative uncertainties 
are at the 1 $\%$ level, similar to full NNLO determinations \cite{eweak}.
\section{Conclusion}
\begin{figure}[h!]
  \unitlength1mm
  \begin{picture}(75,95)
    \put(-2,-1){\includegraphics{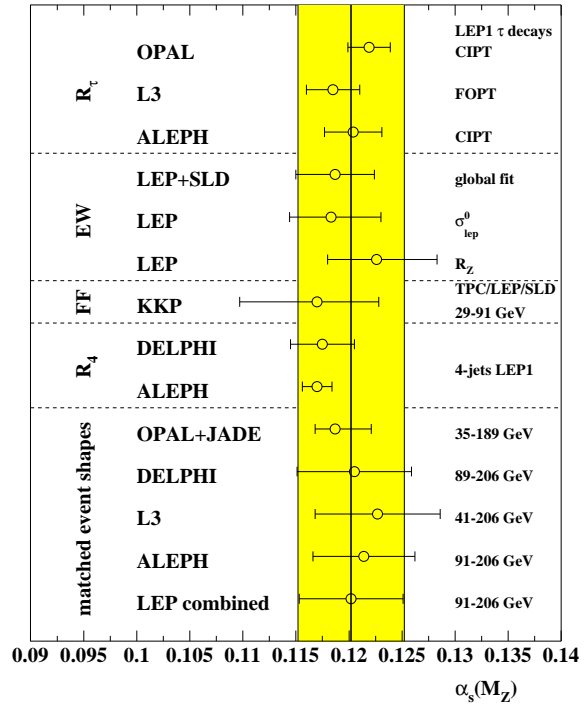}}
  \end{picture}
  \caption{Summary of measurements of $\alpha_s(\mz)$ at LEP using different methods 
and observables.}
\label{fig:chart}
\end{figure}

A wealth of measurements of hadronic observables has been provided in 11 years of data 
taking at the LEP collider. These measurements allowed the LEP collaborations to perform 
detailed tests of perturbative QCD and determinations of the fundamental parameters. A collection 
of the most important results is shown in fig.~\ref{fig:chart}.
 All collaborations have provided 
measurements of $\alpha_s$ from event-shape distributions and combined individual results 
from various variables at energies from $\mz$ to 209 GeV. The systematic uncertainties are 
dominated by missing higher orders and in order to match the experimental accuracy 
of 1 $\%$ NNLO calculations are required. Event-shapes have further been investigated in 
the context of power-law corrections. Simultaneous fits  
have been carried out to determine the non-perturbative parameter $\alpha_0$ and $\alpha_s$. 
While the qualitative description of the data is good, the quantitative interpretation is 
not unambiguous since the value of $\alpha_s$ with power corrections is significantly lower 
than with Monte Carlo corrections for hadronisation. Furthermore, the predicted universality 
of $\alpha_0$ is verified at the level of 20$\%$ of its precision only. In particular the 
jet broadening variables need to be further investigated.

Complete NLO calculations for 4-jet production improved significantly the determinations 
of the QCD colour factors, which are found to be in good agreement with SU(3). New measurements 
of $\alpha_s$ from the 4-jet rate using these calculations have been presented. The precision 
of this method is very good and yields a 1 $\%$ uncertainty for missing higher orders. The 
total uncertainty is of 3 $\%$, dominated by model uncertainties. The 4-jet method is comparable 
in precision with determinations using the fully inclusive observables $R_Z$ and $R_\tau$, for 
which complete NNLO calculations are available.

\end{document}